\documentclass[conference]{IEEEtran}  
\IEEEoverridecommandlockouts                       
\usepackage{cite}
\usepackage{amsmath,amssymb,amsfonts}
\usepackage{algorithmic}
\usepackage{graphicx}
\usepackage{textcomp}
\usepackage{mathtools}
\usepackage{xcolor}
\usepackage{wrapfig}
\usepackage{mathtools}
\usepackage{graphicx}
\usepackage{color}
\usepackage{balance}
\usepackage{verbatim}
\usepackage[normalem]{ulem}
\usepackage{dsfont}
\usepackage[T1]{fontenc}
\usepackage{listings}
\usepackage{url}
\usepackage{authblk}
\usepackage[utf8]{inputenc}
\graphicspath{ {./images/} }
\usepackage{acronym}
\usepackage{cite}
\usepackage{lettrine}
\usepackage{subfigure}
\usepackage{enumitem}
\usepackage{array}

\def\BibTeX{{\rm B\kern-.05em{\sc i\kern-.025em b}\kern-.08em
    T\kern-.1667em\lower.7ex\hbox{E}\kern-.125emX}}
    \newcolumntype{M}[1]{>{\centering\arraybackslash}m{#1}}
\begin{document}
% \title{\fontsize{22.8}{27.6}\selectfont Efficient Handover Mechanism for Cache-Enabled Vehicular Networks}

% \title{\fontsize{22.8}{27.6}\selectfont Handover Rate and Throughput Characterization in UAV Networks}
% \title{\fontsize{22.8}{27.6}\selectfont Measurement selection strategies in ToA based Indoor UE position estimation in NLoS environments}

\title{\fontsize{22.8}{27.6}\selectfont Measurement Selection Strategies for Position Estimation in Indoor Environments}
% \author[1] {Neetu R.R}
% \author[2] {Gourab Ghatak}
% \author[1] {Vivek Ashok Bohara}
% \author[1]{Anand Srivastava}
% \affil[1]{Department of Electronics and Communication Engineering, IIIT Delhi, India}
% \affil[2]{Department of Electrical Engineering, IIT Delhi, India}
\author{Neetu R R, Shrihari Vasudevan and Ranjani Hosakere Gireesha \\
Ericsson R\&D, India\\
Email: \{neetu.rr,  shrihari.vasudevan, ranjani.h.g\}@ericsson.com}

% \thanks{Neetu R.R, Vivek Ashok Bohara and Anand Srivastava are with the Department of Electronics and Communication Engineering, IIIT-Delhi, India. (email: \{neetur, vivek.b, anand\} @iiitd.ac.in) and Gourab Ghatak is with the Department of Electrical Engineering, IIT Delhi, India (email: gghatak@ee.iitd.ac.in).}

% \thanks{The authors are with the Department of Electronics and Communication Engineering, IIIT-Delhi, India. email: \{neetur, gourab.ghatak, anand, vivek.b\} @iiitd.ac.in.}}

\acrodef{AP}[AP]{access point}
\acrodef{LoS}[LoS]{line-of-sight}
\acrodef{IoT}[IoT]{internet-of-things}
\acrodef{NLoS}[NLoS]{Non-line-of-sight}
\acrodef{SINR}[SINR]{signal-to-interference-plus-noise ratio}
\acrodef{QoS}[QoS]{quality of service}
\acrodef{SNR}[SNR]{signal-to- noise ratio}
\acrodef{QoS}[QoS]{quality of service}
\acrodef{CRLB}[CRLB]{cramér-rao lower bound}
\acrodef{UE}[UE]{user equipment}
\acrodef{RSRP}[RSRP]{reference signal received power}
\acrodef{RSS}[RSS]{received signal strength}
\acrodef{RF}[RF]{radio frequency}
\acrodef{PE}[PE]{position error}
\acrodef{CSI}[CSI]{Channel State Information}
\acrodef{3-D}[3-D]{three-dimensional}
\acrodef{2-D}[2-D]{two-dimensional}
\acrodef{1-D}[1-D]{one-dimensional}
\acrodef{5G}[5G]{fifth-generation}
\acrodef{ToA}[ToA]{time-of-arrival}
\acrodef{RT}[RT]{ray-tracing}
\maketitle

\begin{abstract}
% Accurate position estimation in indoor and dense environments is challenging due to occlusion-induced \ac{NLoS} propagation, which introduces significant delay in time-based measurements. In this paper, we propose novel measurement selection strategies for \ac{UE} position estimation that leverage environment-specific propagation characteristics obtained through ray tracing. Using this information, we propose an \ac{AP} neighborhood estimation using delay characteristics of the environment. This neighborhood is used as an apriori information to explore
% % that estimates an apriori knowledge of the neighborhood of the \acp{AP} based on the delay characteristics of the environment. 
% % This neighborhood information is used to explore 
% multiple measurement selection strategies. 
% % we develop reference-\ac{AP}–based framework that provides apriori knowledge of reliable (LoS dominant or less delay) measurements. Within this framework, we investigate and compare multiple strategies, 
% Specifically, we propose a deterministic selection, AP neighborhood–based cardinality selection, intersection and union of AP neighborhood-based selection. Experiments demonstrate the proposed AP neighborhood-based measurement selection strategies significantly improve positioning accuracy 
% % by mitigating the impact of 
% in \ac{NLoS} environments.
Time-based indoor positioning techniques rely on multiple \acp{AP} and measurements between the \ac{UE} and the APs. In dense indoor environments, occlusion-induced \ac{NLoS} propagation introduces significant delays in these measurements, thereby degrading position estimation accuracy. To address this challenge, this paper proposes measurement selection strategies to improve position estimation accuracy. A \ac{RT} simulator is employed to characterize the propagation environment and derive \ac{AP} neighborhood information, which is subsequently used to design and evaluate different measurement selection strategies. The approaches explored include AP neighborhood–based cardinality selection, intersection and union of measurements from AP neighborhoods and fixed measurement selection. Experiments demonstrate the efficacy of the proposed measurement selection strategies in environments under significant \ac{NLoS} conditions. 
\end{abstract}


\section{Introduction}
Accurate position estimation is a fundamental requirement for a wide range of applications, including wireless communications, autonomous systems, and location-based services~\cite{pos_sensing}.
%~\cite{positioning,pos_sensing}. 
In particular, emerging wireless networks such as \ac{5G} and beyond, rely on precise \ac{UE} (such as \ac{IoT} devices, smartphones and tablets) positioning to enable advanced functionalities, such as beam management, resource allocation, and context-aware services~\cite{Beam_form}. Time-based position estimation techniques using \ac{ToA} and timing advance (TA) measurements~\cite{ML_ToA}, are widely used due to their strong physical interpretation and compatibility with existing wireless infrastructure. Under ideal \ac{LoS} conditions, these measurements provide accurate distance estimates that can be used to determine the user location using geometric or optimization-based approaches~\cite{LoS_ana}. However, measurements from \ac{NLoS}~\cite{NLoS_loc} conditions pose significant challenges for accurate position estimation due to the introduction of multipath induced propagation delays in the observed measurements.
% it has been well established that in realistic indoor and dense urban environments~\cite{5},
% % ~\cite{5,6}, 
% \ac{ToA} measurements from \ac{NLoS} conditions results in large variations of measured propagation delay, leading to biased distance estimates and significant positioning errors. 
The classification of measurements as being LoS/NLoS, in real-time, is non-trivial because of the unavailability of the channel information, inability to store the channel data, weak signals or the lack of labelled datasets.

% Therefore, selecting a reliable (less delay or LoS dominan subset of ToA measurements for position estimation is critical to mitigating the impact of NLoS measurements and achieving acccurate indoor positioning.
In this work, we focus on ToA-based position estimation in indoor environments. Recent advances in \ac{RT} simulators enable the generation of environment-specific and physics-guided propagation characteristics.
% ~\cite{RT_AoA}. 
We use Sionna \ac{RT} simulator~\cite{AoA_loc}  to generate uplink/downlink paths between the transmitters and receivers, and capture the corresponding propagation delays for any given environment; ToA measurements are derived from these. 
We propose to improve position estimation accuracy in dense indoor environments by selecting a subset of the ToA measurements that will likely have been produced under LoS conditions.
% provide ToA measurements  and can be used as reliable measurements (e.g., LoS-dominant or less affected by excess delay) for positioning~\cite{8}. 
% In particular, we observe that it is possible to distinguish measurements that are likely to be LoS-dominant from those affected by NLoS propagation by analyzing the \ac{ToA} measurements derived from propagation delays. 

\textbf{Related work}: 
Several studies
%\cite{RT_fingerprint,DL_RT,AoA_loc} and 
\cite{Opt_ToA,ToARSS_mea,opt_sensor_ToA,TOA_RSS_AOA,ToA_TDoA} have proposed diverse frameworks for enhancing the accuracy of \ac{UE} position estimation. The authors in \cite{Opt_ToA} address the challenge of optimally placing \ac{ToA} sensors to localize multiple targets simultaneously. It introduces a novel optimization model that incorporates shared sensors and non-shared sensors dedicated to specific targets. The authors propose a hybrid solution that combines analytical derivations for sensor placement with a numerical optimization algorithm to solve complex scenarios involving various practical constraints. In \cite{ToARSS_mea}, the authors develop optimal sensor selection strategies for \ac{3-D} wireless positioning using hybrid \ac{ToA} and \ac{RSS} measurements. The authors propose two greedy algorithms for dynamic sensor selection when an approximate target location is known, achieving significantly reduced computational complexity. When the target location is unknown, they introduce three optimization techniques, such as iterative convex optimization, difference of convex functions programming, and discrete monotonic optimization to reliably minimize the worst-case \ac{CRLB}. The authors in \cite{opt_sensor_ToA} introduce a robust sensor placement framework for source localization using either \ac{ToA} or \ac{RSS} measurements. To address the challenge of an unknown source position, the authors minimize either the average or worst-case \ac{CRLB} over a grid of potential source locations within a specific region. They utilize a block majorization-minimization  optimization algorithm to solve this non-convex optimization problem. The authors in \cite{TOA_RSS_AOA} propose an optimal sensor placement methodology for source localization by fusing \ac{ToA}, \ac{RSS}, and Angle of Arrival (AoA) measurements. To maximize accuracy, the authors formulate a design problem based on minimizing the trace of the CRLB. They introduce a majorization-minimization algorithm and transform it into an equivalent saddle-point problem using an auxiliary variable.

Existing works have proposed various sensor/AP selection strategies for position estimation. However, they do not consider measurement selection based on apriori information about AP neighborhoods. Such information captures environment-specific propagation characteristics and can be exploited to select less-delayed measurements, thereby improving position estimation accuracy. Motivated by this, the paper investigates ToA measurement selection strategies that leverage \ac{AP} neighborhood estimates derived from ray-tracing–based propagation information.

% we propose a ToA-based positioning framework that utilizes ray-tracing–derived propagation information to design effective measurement selection strategies, thereby enhancing positioning accuracy in indoor environments.

The main contributions of this paper are:
\begin{itemize}
    % \item  We construct a realistic three-dimensional indoor propagation environment by converting a \ac{2-D} floor plan into a \ac{3-D} scene, enabling accurate characterization of propagation delays and interaction effects between transmitters and receivers using ray tracing.
    
    % \item We develop a radio map-assisted \ac{ToA}-based user positioning framework that accounts for the impact of \ac{NLoS} propagation. The user position is estimated by formulating and solving a nonlinear least squares problem to mitigate the effect of measurement noise and propagation-induced errors.

    \item We demonstrate that selecting an appropriate subset of ToA measurements improves position estimation accuracy in mixed \ac{LoS} and \ac{NLoS} scenarios.

    % \item We propose a \ac{RT} based AP neighborhood estimation as an apriori information to ToA measurement selection strategies.

    \item Towards doing position estimation using a subset of the ToA measurements, we leverage an \ac{RT} simulator to characterize environment-specific propagation and derive AP neighborhood information.
    
    \item We propose four ToA measurement selection strategies: (i) neighbourhood-based
selection, (ii) intersection-based selection, (iii) union-based selection, and (iv)  cardinality-based selection

    % minimize position estimation error, which, to the best of our knowledge, has not been previously explored. We leverage ray-tracing tools to model propagation paths between \acp{AP} and \ac{IoT} devices, and extract \ac{ToA} measurements based on propagation delay characteristics.
    % % The network model and corresponding non-linear least square algorithm is presented.

    % \item  We introduce an elbow-based method to identify the optimal number of reliable measurements by distinguishing \ac{LoS}-dominant measurements from \ac{NLoS}-affected measurements. Due to variations in AP-IoT-device propagation conditions, we investigate and compare multiple measurement selection strategies, including deterministic selection, cardinality-based selection, intersection-based selection, and union-based selection.
    
    % \item Simulation results indicate that the proposed measurement selection approaches significantly outperform the all-measurement selection baseline in terms of position estimation accuracy under \ac{NLoS} propagation conditions.
\end{itemize}

\section{Indoor NLoS Network Model}
% We detail the position estimation formulation (\ref{ssec:formulate}) followed by an overview of scenario simulation (\ref{ssec:scene_simulate})
% \subsection{Formulation}
% \label{ssec:formulate}
Consider an indoor wireless network consisting of $M$ 
\acp{AP}. The locations of APs are known apriori and denoted by
$\mathbf{p}_\mathrm{m}= [ x_\mathrm{m},y_\mathrm{m},z_\mathrm{m}]$ $\in$ $\mathbb{R}^3$, $m=1 \dots M$.
The positions of \ac{UE} are unknown and denoted by $\mathbf{p}_\mathrm{u}= [ x_\mathrm{u},y_\mathrm{u},z_\mathrm{u}]$ $\in$ $\mathbb{R}^3$. Due to the presence of occlusions in the indoor environment, propagation paths between the UE and APs tend to include both LoS and NLoS paths. These paths create delays, the delay corresponding to the dominant path (least delayed path) between the $m^{\textrm{th}}$ \ac{AP} and the UE is denoted as $\tau_\mathrm{m}$. Geometrically, the user position is estimated via multilateration~\cite{multilateral} corresponding to the intersection region of a set of $M$ circles (in \ac{2-D}; spheres in \ac{3-D}). In LoS conditions, the radius corresponds to derived Euclidean distance\footnote{The Euclidean distances can be derived from time-based measurements}, $|| \mathbf{p}_\mathrm{u} - \mathbf{p}_\mathrm{m}||_2$; the region of intersection includes $\mathbf{p}_m$, and $\mathbf{p}_\mathrm{u}$ can be estimated by solving equations, quantifying the pairwise Euclidean distance between the \ac{UE} and every AP. 

% Under NLoS conditions, 
Due to the presence of occlusions, the measured propagation delays do not directly correspond to the Euclidean distance between the AP and the device, i.e., $|| \mathbf{p}_\mathrm{u} - \mathbf{p}_\mathrm{m}||_2 \neq c \tau_m$.  The objective is to minimize the discrepancy between the measured distances and the distances indicated by a candidate user position. The residual corresponding to the $m^{\textrm{th}}$ \ac{AP} is defined as
\begin{equation}
    r_\mathrm{m}= || \mathbf{p}_\mathrm{u} - \mathbf{p}_\mathrm{m}||_2 - c (\tau_{\mathrm{fm}}+ \tau_\mathrm{m}),
\end{equation}
where $c$ is the speed of light and $\tau_{\mathrm{fm}}$ is the time of flight i.e., the time
at which the pilot signal is sent by the UE. 
The \ac{ToA} at the $m^{\mathrm{th}}$ \ac{AP} is given as $\mu_\mathrm{m} = \tau_{\mathrm{fm}} + \tau_\mathrm{m}$.
% where $d_m= \tau_m c$ is the distance between the $m^{\textrm{th}}$ AP and UE in the presence of measurement noise and propagation effects such as blockages and multipath. 
In this work, we assume perfect time synchronization between \acp{UE} and \acp{AP}. Therefore, without loss of generality, we set $\tau_{fm}=0$, and $\mu_m = \tau_m$. Considering the residuals for all the $M$ \acp{AP}, the UE position is estimated as
\begin{equation}
\hat{\mathbf{p}}_\mathrm{u} =
\arg \min_{\mathbf{p}_\mathrm{u}} \sum_{m=1}^{M} \left(
\| \mathbf{p}_\mathrm{u} - \mathbf{p}_\mathrm{m} \|_2 - c \mu_\mathrm{m}\right)^2 .
\label{nnls_1}
\end{equation}
This is a nonlinear least-squares optimization formulation.
% ~\cite{Adaptive_NLLS}; . 
The solution employs an iterative trust-region based algorithm~\cite{NLLS}. From an initial estimate, the method minimizes the residual error between measured and predicted distances through successive linear approximations until convergence.

The \ac{PE}, $e$, is defined as
    $e= || \hat{\mathbf{p}}_\mathrm{u} - \mathbf{p}_\mathrm{u} ||_2$.

\subsection{Indoor environment simulation in an RT simulator}
\label{ssec:scene_simulate}
\begin{figure}[h!]
\centering
\includegraphics[width=0.55\linewidth, height=0.6\linewidth]{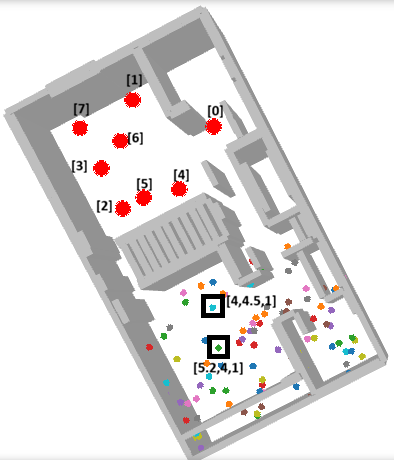}
\caption{A 3-D scene illustrating eight APs, represented as large, numbered red markers, and uniformly distributed \ac{UE} locations, indicated by small multicolored markers. In this paper, two boxed UE locations are selected as reference points for analysis; the first UE location  [5.2,4,1] represents a weak-\ac{NLoS} environment, and the other [4,4.5,1] represents a strong-\ac{NLoS} environment.}
\label{Network_mod}
\end{figure}
Fig.~\ref{Network_mod} illustrates a representative weak NLoS and strong NLoS scenario. The weak and strong \ac{NLoS} signals are distinguished based on propagation characteristics such as the number of reflections and the associated path attenuation before reaching the \ac{UE}.
% The red dots are the \acp{AP} and the UEs (multicolored) are randomly sampled from uniform distribution. 
The UEs are placed in a region occluded from the APs to simulate NLoS conditions. The proposed approach, detailed in \ref{sec:measurement_selection}, however, is generic and not restricted to this specific layout. Advances in \ac{3-D} sensing and software tools have made detailed \ac{3-D} indoor representations increasingly accessible. Where 3D representations for indoor environments are not readily available but 2D floor plans are, it is possible to extrude a 2D floor plan into a \ac{3-D} model with coarse material properties. In this work, a \ac{2-D} floor plan is used to construct a \ac{3-D} representation of the indoor environment, explicitly incorporating structural elements such as walls that act as propagation blockages.
% A two-dimensional floor plan is used to construct a three-dimensional representation of the indoor environment, which explicitly includes structural elements such as walls that act as propagation blockages. 
These blockages influence the propagation characteristics by reflecting and/or completely obstructing the radio signal. The captured propagation delays are indicative of environmental blockages and multipath effects.

We use ray-tracing tools~\cite{RT_fingerpr} that explicitly account for geometric layout and environmental blockages~\cite{ToA_AoA} to generate environment-specific, physically accurate channel realizations for the considered indoor scene and AP positions.
For each AP-\ac{UE} pair, the ray-tracing tool provides path attenuation and propagation delays. We consider the minimum-delay path for each AP-UE pair to derive \ac{ToA} measurements. 

However, utilizing measurements from all $M$ \acp{AP} introduces additional multipath-induced delays, which degrade position estimation accuracy, particularly under multipath conditions. To address this, we propose selecting a subset of \ac{ToA} measurements that are more likely to originate from \ac{LoS} or weak \ac{NLoS} paths. The challenge lies in identifying measurements that improve position estimation while maintaining low complexity for practical deployment. To achieve this, we determine an AP neighborhood for each AP. Using this information, we develop different measurement selection strategies to identify suitable subsets for position estimation. In the next section, we describe these approaches in detail.

\section{Proposed Approach}
\label{sec:measurement_selection}

In this section, we propose measurement selection strategies to improve position estimation accuracy. A two-stage process is presented- (i) Estimation of AP neighborhoods using path delays derived from ray tracing, (ii) measurement selection based on the neighborhood of the nearest AP, determined by the minimum ToA.

\subsection{AP Neighborhood Estimation}
\subsubsection{AP Neighborhood definition}
\label{elbow-mension}
\begin{figure*}[t]
    \centering    \subfigure[\label{Ref_AP_fig}]{\includegraphics[width=0.24\linewidth, height=0.18\linewidth]{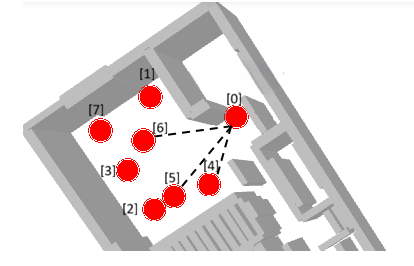}}
    \subfigure[\label{LK_tab}]{\includegraphics[width=0.24\linewidth, height=0.18\linewidth]{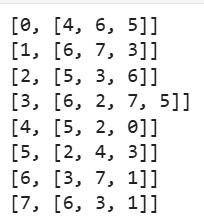}}    \subfigure[\label{elbow_1}]{\includegraphics[width=0.24\linewidth, height=0.18\linewidth]{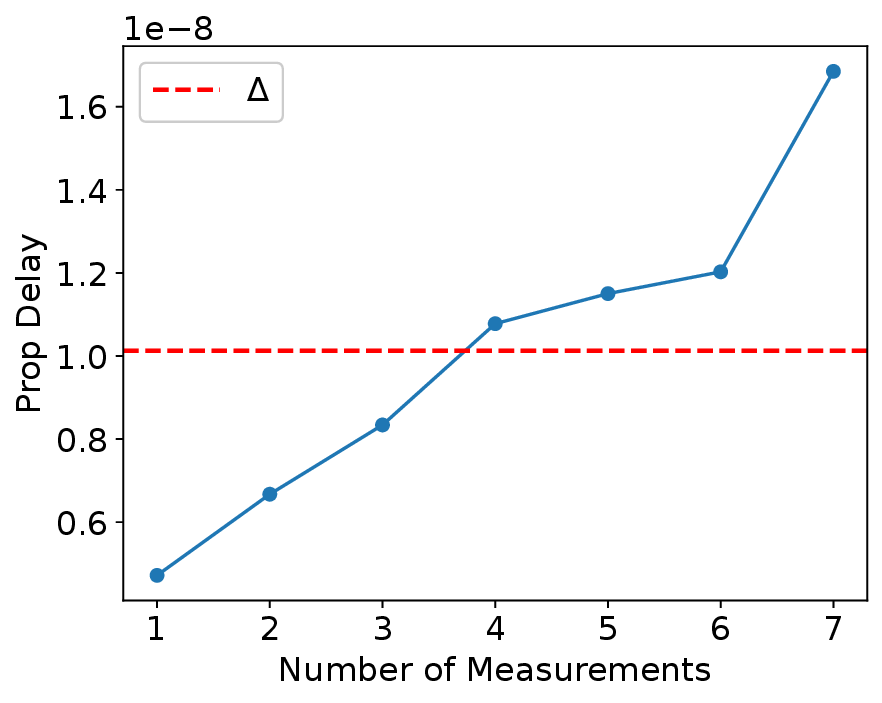}}
    \caption{(a) A layout with the AP-0 and its neighborhood represented by dotted lines (b) Look-up table for single transmitter reference AP (c) Elbow curve for the AP-0 with 3 optimal measurements below the threshold }
\end{figure*}

Given a floorplan and/or 3D scene along with $\mathbf{p_m}$, we employ ray-tracing to obtain path delay estimates by sequentially treating each $m^{\mathrm{th}}$ AP as a transmitter and all remaining APs as receivers. This process yields ToA measurements between all AP pairs, which are subsequently used to define the neighborhood of each AP, as given in Fig.~\ref{Ref_AP_fig}. 
% Let the set of ToA measurements obtained from $M-1$ APs for determining the neighborhood be denoted by $\{\mu_{(1)},\mu_{(2)}, \dots, \mu_{(M-1)}\}$. 
Let $B_{\mathrm{m}}$ denote the set of neighboring APs of the $m^{\text{th}}$ AP, where $
B_{\mathrm{m}} \subseteq \{1,2,\ldots,M\} \setminus \{m\}$. 
% The corresponding ToA measurements $\{\mu_i^{(m)} \mid i \in B_\mathrm{m}\}$ are sorted in ascending order. 
${B}_{\mathrm{m}}$ is estimated\footnote{Note that this differs from a Euclidean distance–based neighborhood, as the ToA-based neighborhood inherently accounts for occlusions, whereas the Euclidean distance–based approach does not} based on propagation characteristics that influence the \ac{ToA} measurements of $m^{\mathrm{th}}$ AP.  ${B}_{\mathrm{m}}$ is an indicator of apriori knowledge of the neighboring APs in the given environment. From the set ${B}_{\mathrm{m}}$, the number of measurements for position estimation is determined using the Elbow method applied to the corresponding propagation delay values.

% \subsubsection{Elbow Method}
% \label{elbow_meth}
The Elbow method is a simple change-point detection technique used to determine the number of ToA measurements to be selected for estimating the UE position, denoted by $k$. It identifies a point at which the propagation delays exhibit a pronounced increase, indicating a transition beyond a specified threshold. This method is commonly applied in K-means clustering to determine the number of clusters~\cite{K_means}. In this work, the ToA measurements are first sorted in ascending order and plotted against their index. The elbow point is then identified as the location where a sudden shift in slope occurs, separating measurements with lower delays from those affected by higher delays.

\subsubsection{AP Neighborhood estimation using Elbow method}
For a given AP $m$, let the sorted ToA measurements corresponding to the neighboring APs in $B_{\mathrm{m}}$ be given by
\begin{equation}
\left\{ \mu_{(1)}^{(m)}, \mu_{(2)}^{(m)}, \dots, \mu_{(|B_{\mathrm{m}}|)}^{(m)} \right\},
\label{sort_toa}
\end{equation}
where $\mu_{(i)}^{(m)}$ denotes the ToA measurement between the $m^{\text{th}}$ AP and its neighboring AP $i$.
% These ToA measurements are sorted as
% \begin{equation}
% \left\{ \mu_{(1)}^{(m)}, \mu_{(2)}^{(m)}, \dots, \mu_{(|B_m|)}^{(m)} \right\},
% \end{equation}
$\mu_{(1)}^{(m)}$ represents the smallest ToA value of the $m^{\mathrm{th}}$ AP and $|B_{\mathrm{m}}| \leq  M-1$.
The differences of these delays between these $M-1$ \acp{AP} are
\begin{equation}
\delta_\mathrm{i}^{(m)} =
\mu_{(i+1)}^{(m)} -
\mu_{(i)}^{(m)},
\quad i = 1, 2, \dots, M-2.
\end{equation}
From these delay differences $\delta_i^{(m)}$, we identify the index at which the propagation delay exhibits a significant increase using the Elbow method described above. 
Therefore, the number of measurements $k$ considered for the $m^{\mathrm{th}}$ AP is given by
\begin{equation}
k^{(m)} = \arg \min_{i} \left\{ i \;:\; \mathbb{I}\!\left(\delta_i^{(m)} > \Delta \right) = 1 \right\},
% \Delta_i^{(r)}
\label{eq_delta}
\end{equation}
where $k^{(m)} \subseteq \{3,4, \dots , M\}$ denotes the selected number of ToA measurements. $\Delta$ is the sample mean of the difference of delays between $M-1$ \acp{AP}, given as
\begin{equation}
\Delta = \frac{1}{M-2} \sum_{i=1}^{M-2} \delta_\mathrm{i}^{(m)}.
\end{equation}
Using the elbow method, a look-up table is constructed for each AP, as shown in Fig.~\ref{LK_tab}. An example elbow curve corresponding to $k^{(m)}$ is illustrated in Fig.~\ref{elbow_1}. The lower bound of $k^{(m)}$ is set to 3, because number of
ToA measurements less than 3 will lead to inaccurate estimations.

Let $\{j_{(1)}, j_{(2)}, \dots, j_{(|B_m|)}\}$ be an ordering of the indices in $B_m$ such that
\begin{equation}
\mu_{j_{(1)}}^{(m)} \leq \mu_{j_{(2)}}^{(m)} \leq \cdots \leq \mu_{j_{(|B_m|)}}^{(m)}.    
\end{equation}
Therefore, after applying the Elbow method, the AP neighborhood of the $m^{\mathrm{th}}$ AP is obtained as.
\begin{equation}
N_m = \{ j_{(1)}, j_{(2)}, \dots, j_{(k^{(m)})} \}, \quad N_m \subseteq B_m,
\end{equation}
which contains the indices of the AP neighborhood corresponding to the $k^{(m)}$ smallest ToA measurements with respect to the $m^{\mathrm{th}}$ AP. 
Extending this across all APs, these measurements represent the smallest propagation delays and are thus suitable for accurate user position estimation. Using this information, different measurement selection strategies for position estimation are presented below. 

\subsection{Measurement Selection Strategies}
We consider a scenario in which the \ac{UE} location is unknown and ToA measurements are available from $M$ \acp{AP}. The reference AP is selected as the AP corresponding to the minimum ToA with respect to the UE; without loss of generality, the UE is assumed to be associated with this AP. Based on prior AP neighborhood estimation, the neighborhood of the selected reference AP is known. However, directly utilizing all ToA measurements from these APs may include both \ac{LoS} and \ac{NLoS} components, leading to increased \ac{PE}. To address this, we propose measurement selection strategies that exploit the AP neighborhood information to improve positioning accuracy.

Given the AP neighborhood information, four measurement selection strategies are proposed: (1) neighborhood-based selection, (2) intersection-based selection using two reference \acp{AP}, (3) union-based selection using two reference \acp{AP} and (4) cardinality-based selection. Further, two measurement selection strategies without any AP neighborhood information are proposed: (1) fixed measurement selection, (2) fixed elbow-based measurement selection.

% Here, we propose:
% \begin{itemize}
%     \item four selection strategies which use AP neighborhood estimation to improve positioning accuracy. They are (1) $N_{\mathrm{m}}$ neighborhood based selection (2) $N_{\mathrm{m}}$ intersection-based selection using two reference \acp{AP} (3) $N_{\mathrm{m}}$ union-based selection using two reference \acp{AP} and (4) $N_{\mathrm{m}}$ cardinality-based selection,.
%     \item two measurement selection strategies without any AP neighborhood apriori information - (5) deterministic approach and (6) deterministic elbow approach
% \end{itemize}.

% We consider a reference AP defined as the AP corresponding to the minimum ToA with respect to the UE. Without loss of generality, we assume that the UE associates with this AP. 
Let the ToA measurements obtained at the $M$ APs with respect to the UE be denoted by $
\{\mu_1, \mu_2, \dots, \mu_M\} $. The reference AP, denoted by $r$, is defined as the AP corresponding to the minimum ToA. 

\subsubsection{Neighborhood-based selection} The AP neighborhood-based approach uses the look-up table shown in Fig.~\ref{LK_tab} to identify the set of neighboring APs corresponding to the reference AP $r$. Only the ToA measurements associated with this neighborhood are considered.
For a given reference AP $r$, let $N_\mathrm{r}$ denote the corresponding set of neighboring AP indices, where $r \notin N_r$. The associated ToA measurements considered for the UE position estimation are given by
\begin{equation}
\left\{ \mu_r \right\} \cup \left\{ \mu_i \mid i \in N_r \right\},
\end{equation}
where $\mu_r$ denotes the ToA corresponding to the reference AP, and $\mu_i$ denotes the ToA measurements of neighboring APs.
% $i \in N_\mathrm{r}$.

However, AP neighborhood selection based on one reference AP lacks UE-centric perspective, as some APs within a given neighborhood may exhibit strong \ac{NLoS} conditions with respect to the UE. To mitigate this mismatch and preferentially select \ac{LoS} or weak \ac{NLoS} measurements, we extend the single reference AP–based neighborhood selection to approaches that utilize the set union and set intersection of multiple AP neighborhoods.
% Its weakness is that it lacks a UE perspective of AP neighborhood => some of the AP's neighbors may not be truly LoS for the UE.
% To more completely capture the impact of \ac{NLoS} measurements, we extend the single reference AP–based neighborhood selection to approaches that utilize the set union and set intersection of multiple AP neighborhoods. 
Specifically, these methods combine the neighborhood-based measurements corresponding to two reference APs, defined as the APs associated with the two smallest ToA values.

% Next, we consider intersection- and union-based selection methods, which combine the AP neighborhood measurements corresponding to the two nearest APs with respect to the UE (i.e., the APs associated with the two smallest ToA values). Such a combination is expected to improve robustness against blockages by incorporating the whole neighborhood information.

\subsubsection{Intersection-based selection}
% In this method, we consider the common neighborhood measurements by taking the intersection of the AP neighborhood sets corresponding to the reference AP and the AP with the second smallest ToA with respect to the UE. 
This approach selects ToA measurements corresponding to the intersection of the neighborhood sets of the two reference APs. Specifically, let $N_r$ and $N_s$ denote the neighborhood sets obtained from the look-up table for the reference AP $r$ and the second nearest AP $s$, respectively. 
The selected ToA measurements are given by
\begin{equation}
\left\{ \mu_r, \mu_s \right\} \cup \left\{ \mu_i \mid i \in N_r \cap N_s \right\}.
\end{equation}
These ToA measurements correspond to APs that are common to the neighborhood sets of both reference APs, thereby enhancing the likelihood of LoS or weak NLoS measurements for UE position estimation.

\subsubsection{Union-based selection} In this method, the selected set consists of ToA measurements corresponding to the union of the neighborhood sets of the two reference APs.  Specifically, all measurements corresponding to APs that belong to the neighborhood of either reference AP are included in the selected set. To refine this set and limit the impact of delayed measurements, the elbow method is applied to the union set to determine an appropriate subset for position estimation.
The corresponding ToA measurements are given by
\begin{equation}
\left\{ \mu_r, \mu_s \right\} \cup \left\{ \mu_i \mid i \in N_r \cup N_s \right\}.
\end{equation}
These ToA measurements are sorted in ascending order, and the elbow method is applied to determine the optimal number of measurements. Accordingly, the selected set of ToA measurements is given by
\begin{equation}
\left\{ \mu_{1}, \mu_{2}, \dots, \mu_{k^{(r,s)}} \right\},
\end{equation}
where $k^{(r,s)}$ denotes the number of selected measurements obtained using the elbow method on the union set.

% In order to reduce reliance on explicit neighborhood information, we consider a cardinality-based selection approach, which uses the size of the AP neighborhood to determine the number of ToA measurements to be selected, while choosing the corresponding number of smallest ToA measurements instead of relying on the specific neighborhood information.
 \begin{figure*}
    \centering    \subfigure[\label{rxr_ranloc}]{\includegraphics[width=0.275\linewidth, height=0.26\linewidth]{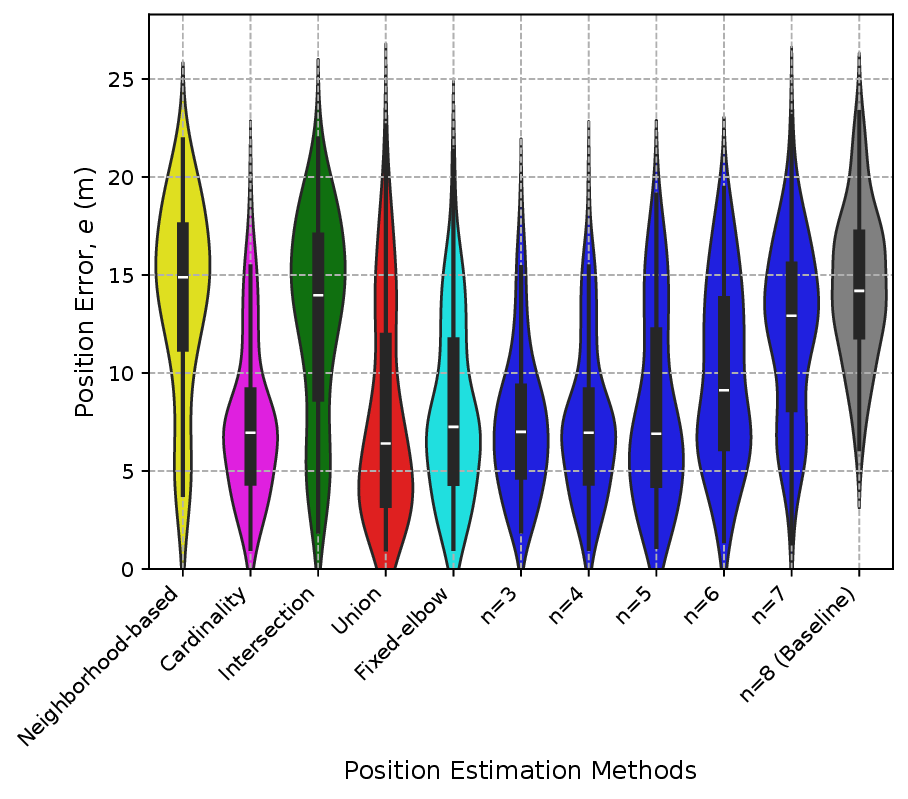}}
    \subfigure[\label{rxr_loc1}]{\includegraphics[width=0.275\linewidth, height=0.26\linewidth]{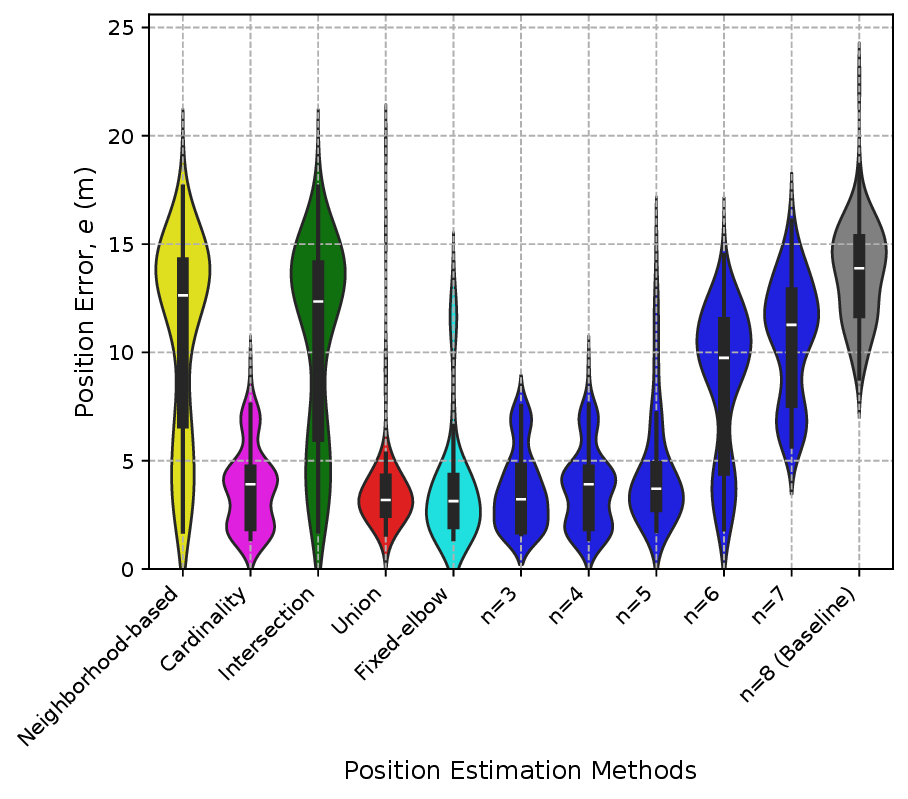}}
    \subfigure[\label{rxr_loc2}]{\includegraphics[width=0.275\linewidth, height=0.26\linewidth]{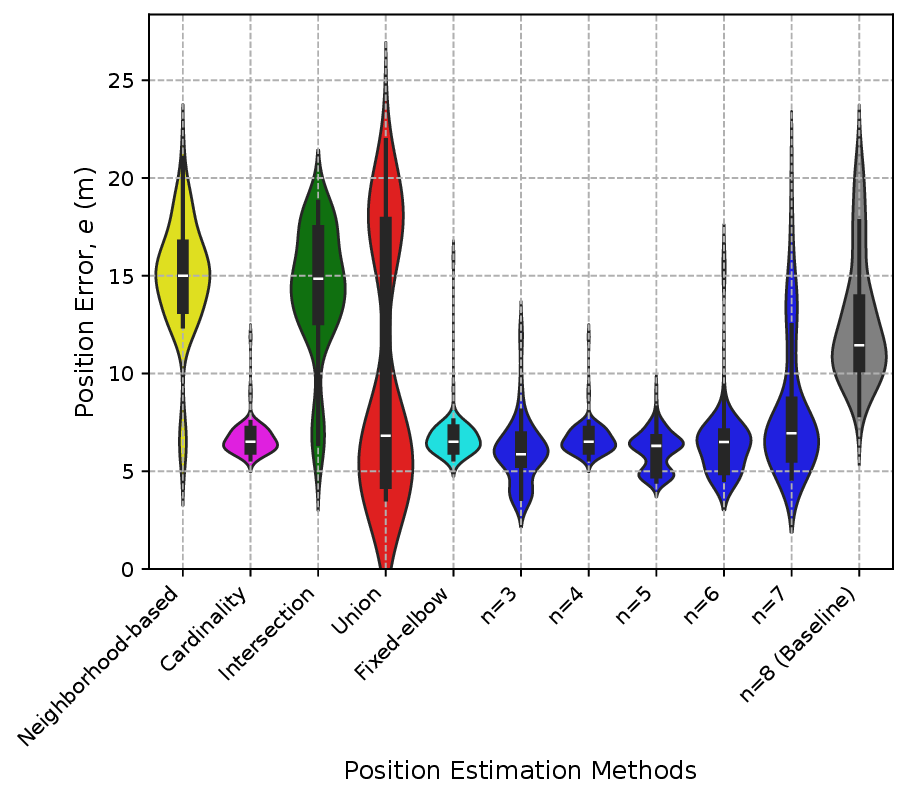}}    \subfigure{\includegraphics[width=0.15\linewidth, height=0.12\linewidth]{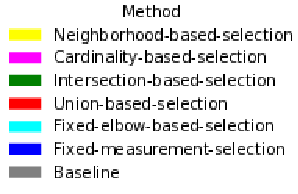}}
    \caption{(a) Comparative plots for different position estimation methods where $\sim$100 UEs are uniformly distributed across the region of interest (b) $\sim$100 UEs uniformly distributed around a reference UE at [5.2,4,1] (c) $\sim$100 UEs uniformly distributed around a reference UE at [4,4.5,1].}
    \label{VP_plots}
\end{figure*}

\subsubsection{Cardinality-based selection} 
To reduce reliance on explicit neighborhood information, we consider a cardinality-based selection approach that utilizes only the size of the AP neighborhood. Let $|N_r|$ denote the number of neighbors associated with the reference AP $r$. The ToA measurements with respect to the UE are sorted in ascending order, and the first $|N_r|$ measurements are selected. Based on the cardinality $|N_r|$, the number of measurements is selected for the reference AP.
% and the first $|N_r|$ measurements from the sorted sequence of ToA measurements are considered. 
Accordingly, the selected measurements are given by
\begin{equation}
\left\{ \mu_r \right\} \cup \left\{
\mu_{1}, \mu_{2}, \dots, \mu_{|N_r|}
\right\}.
\end{equation}

Next, we discuss measurement selection strategies that do not rely on apriori AP neighborhood information; instead, a fixed number $n$ of measurements is selected deterministically. Such deterministic selection of a fixed number of measurements is straightforward to implement and ensures minimal computational complexity.

\subsubsection{Fixed measurement selection}

In this method, we consider $n$ ToA measurements, $(n < M)$, obtained from $M$ APs in a deterministic manner, i.e., $n$ is fixed apriori. First, the ToA measurements are sorted in ascending order, and the first $n$ measurements are selected for position estimation, where $n = \{4,5, \dots, M\}$. In scenarios with dominant LoS conditions, $n$ can be reduced to 3.
% These $n$ measurements correspond to the smallest propagation delays, and are used for user position estimation. 
The selected ToA measurements are given as
\begin{equation}
\left\{ \mu_{1}, \mu_{2}, \dots, \mu_{n} \right\}.
\end{equation}

\subsubsection{Fixed Elbow-based measurement selection}
In this method, the \ac{ToA} measurements from all \acp{AP} with respect to the UE are sorted in ascending order, and the Elbow method is applied to determine the number of measurements $k$ (detailed in \ref{elbow-mension}). Unlike the AP neighborhood–based approach, the Elbow method is applied directly to the complete set of ToA measurements observed at the APs with respect to the UE.

The selected ToA measurements are given as
\begin{equation}
\left\{ \mu_{1}, \mu_{2}, \dots, \mu_{k} \right\},
\end{equation}

\section{Numerical Results and Discussion}
In this section, we perform Monte Carlo simulations on a sample indoor scenario, present and analyze numerical results to discuss the salient features of the model. 
\subsection{Experimental setup}
\label{ssec:expt_setup}
The formulation described in Section 
\ref{sec:measurement_selection} is generic and is applicable in mixed LoS/NLoS and pure LoS scenarios. For experimental purposes, we consider an indoor scenario and we simulate eight \acp{AP} at fixed manually chosen locations.  A total of 100 UEs are uniformly distributed across the region of interest (uniformly distributed across a larger region or small selected locations as given in Fig.~\ref{Network_mod}). We use NVIDIA Sionna ray-tracing tool to generate paths between the APs and UEs. We consider a region such that there are weak and strong NLoS paths. The transmit power of AP is $44~\mathrm{dBm}$, the bandwidth of the system is $100~\mathrm{MHz}$, the carrier frequency is $3.5~\mathrm{GHz}$.
% and the indoor enclosure dimensions are $[0.13,7.31] \times [0.37,13] \times [0,3]~\mathrm{m}$. 
% \textcolor{red}{see, one issue is your room dimensions are off by a factor of 10. typical indoor is ~70x120 m in 3gpp specs with trp at 8m and ue at 1.5m. I don't know to what extent you can regenerate these figures 3 and 4 by increasing the room dimension in accordance by a factor of 10 along both x and y dimension }

As a baseline, we consider the all-measurement-selection strategy~\cite{ToA_RSS_mixed_Vis}, where all the measurements from the  APs are used for position estimation. To solve the NLLS optimization problem in (\ref{nnls_1}), we consider the numerical solution obtained using \texttt{scipy.optimize.least\_squares} library. In the figures below, the baseline approach is represented by the case corresponding to $n=8$, shown in grey.

% \renewcommand{\arraystretch}{1}
% \begin{table}[]
%     \centering
%      \caption{Simulation Parameters}
%     \label{TABLE: table11}
%     \begin{tabular}{|M{3cm}|M{4cm}|}\hline
%     Parameter & Value \\\hline\hline
%     Transmit power of AP & 44 dBm \\\hline
%     Bandwidth of the system &100 MHz\\\hline
%      Carrier frequency & 3.5 Ghz\\\hline
%      Noise density & -174 dBm/Hz\\\hline
%      Room dimensions & [0.13,7.31] $\times$ [0.37,13] $\times$ [0,3] m\\\hline
%      % User locations & [5.2,4,1],  [3.5,4.5,1]\\\hline
%       \end{tabular}
%    \end{table}

% \begin{figure}[htbp]
% \centering
% \includegraphics[width=1\linewidth,height=0.85\linewidth]{Test_det_elb_ran.png}
% \caption{Comparative plots for uniformly distributed UE locations}
% \label{rxr_ranloc}
% \end{figure}

\subsection{\acp{UE} uniformly distributed across a larger region}
Fig.~\ref{rxr_ranloc} presents the distribution of \acp{PE} using a violin plot for the six measurement selection strategies across 100 UE locations.
% as described in Section~\ref{ssec:expt_setup}. 
The proposed selection strategies outperform the baseline approach, which results in higher estimation error. The neighborhood-based and intersection methods exhibit higher median error and larger error spread due to the lack of a UE-centric perspective and the use of smaller measurement subsets, respectively. In contrast, the cardinality- and union-based methods achieve lower median error and reduced variance, as the former leverages the neighborhood cardinality to select measurements, while the latter includes a larger set of likely LoS or weak NLoS measurements. 
% by selecting an optimal number of measurements while maintaining a balance between measurement quality and geometric diversity.
% The deterministic elbow-based selection achieves performance comparable to the cardinality-based method but shows a higher spread in the error distribution due to the adaptive nature of the elbow-based selection leading to the selection of more measurements. 
In fixed measurement selection, increasing the number of measurements beyond a certain point leads to performance degradation due to the inclusion of delay-affected measurements. Empirically, selecting 3–4 measurements yields lower position estimation error. These results demonstrate that the cardinality selection and union-based selection provide robustness and improved positioning accuracy.

\subsection{UEs uniformly distributed around a reference UE}
In order to evaluate the sensitivity and robustness of different measurement selection strategies to small spatial perturbations, we analyze the distribution of \ac{PE} around two reference UE locations as given in Fig.~\ref{Network_mod}. Specifically, in Fig.~\ref{rxr_loc1}, we present a violin plot illustrating the distribution of \ac{PE} for different measurement selection strategies, corresponding to a reference UE located at [5.2,4,1] and approximately 100 receiver positions uniformly distributed within a radius of 0.5 m from the UE. 

\begin{figure}[htbp]
\centering
\includegraphics[width=0.7\linewidth,height=0.5\linewidth]{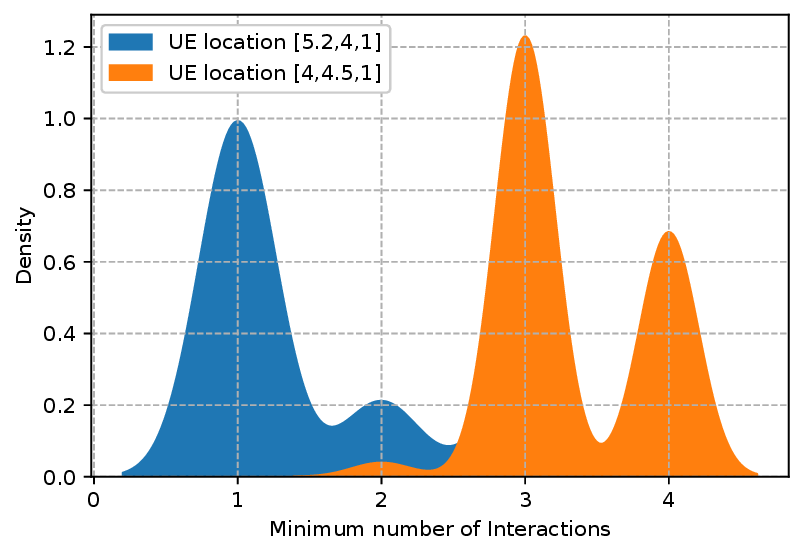}
\caption{Kernel density plot for minimum number of interactions}
\label{INT_CDF}
\end{figure}

We observe that the union-based and fixed elbow-based methods achieve a lower \ac{PE} compared to the other strategies at this UE location. This indicates that combining measurements from multiple reference AP neighborhoods or adaptively selecting measurements based on ToA values results in an effective subset of measurements. Furthermore, the overall \ac{PE} at this UE location is lower than that at the UE location at [4,4.5,1], as shown in Fig.~\ref{rxr_loc2}. The cardinality and fixed elbow-based approach performs the best in this scenario because controlling the cardinality or adaptively identifying an elbow point produces a suitable subset of measurements. Although both UE locations are in \ac{NLoS} conditions with respect to the APs, the difference in positioning performance can be attributed to the number of interactions or reflections experienced by the UE from the APs at these locations, as illustrated in Fig.~\ref{INT_CDF}. 
From Fig.~\ref{INT_CDF}, we observe that the distribution for the second reference UE at [4,4.5,1] (orange curve) is concentrated around 3 to 4 interactions, whereas the other reference UE at [5.2,4,1] (blue curve) has most of its interactions concentrated around 1 to 2. A higher number of interactions indicates that the signals reach the receiver after multiple interactions, which introduces additional propagation delay and increases the likelihood of \ac{NLoS} conditions. These result in higher position estimation error. In contrast, the UE at [5.2,4,1] experiences fewer interactions, meaning more or less-reflected paths, which provide lower \ac{PE}. 

Based on these observations, in indoor NLoS environments with fewer interactions, such as factory floors with some occlusions, selection strategies such as cardinality-based, union-based, or fixed elbow-based methods are suitable, as these measurement selections improve the position accuracy. In contrast, in an NLoS environment having varied interactions within small regions (such as hospitals, universities, office spaces and classrooms), cardinality-based selection shows promise, as it effectively filters high delay measurements, thereby improving positioning accuracy.

\section{conclusion}
In this paper, we proposed an approach to estimate AP-neighborhood using propagation delay information using ray tracing tools for indoor environments. We use this AP-neighborhood to select measurements for position estimation in NLoS scenarios. We proposed six measurement selection strategies. Simulation results demonstrated that the proposed AP-neighborhood based measurement selection strategies (particularly the union-based and cardinality-based methods) significantly improve positioning accuracy compared to a fixed number of measurement selection approach. These results highlight the importance of including environment-specific and propagation-aware measurement selection for indoor positioning. Future work includes evaluating the scalability of the proposed methods in large-scale deployments with dense AP configurations.

\bibliography{references}
\bibliographystyle{IEEEtran}

\end{document}